\documentstyle[12pt,epsf]{article}    

\begin{document}

\baselineskip=25pt

\begin{center}{
{
\Large
A hysteresis model with dipole interaction: one more devil-staircase.
}\\

\vskip 24pt
{\large A.A.Fraerman and M.V. Sapozhnikov}\\
\vskip 12pt
Russian Academy of Science\\
Institute for Physics of Microstructures\\
\vskip -3pt GSP-105, Nizhny Novgorod, 603600, Russia\\
E-mail: msap@ipm.sci-nnov.ru}
\end{center}
\vskip 2mm


\vskip 20mm
\begin{center}
Abstract
\end{center}
Magnetic properties of 2D systems of magnetic nanoobjects (2D regular
lattices of the magnetic nanoparticles or magnetic nanostripes) are
considered.  The analytical calculation of the hysteresis curve of the
system with interaction between nanoobjects is provided. It is shown that
during the magnetization reversal system passes through a number of
metastable states. The kinetic problem of the magnetization reversal was
solved for three models. The following results have been obtained. 1) For
1D system (T=0) with the long-range interaction with the energy
proportional to $r^{-p}$, the staircase-like shape of the magnetization
curve has self-similar character. The nature of the steps is determined
by interplay of the interparticle interaction and coercivity of the
single nanoparticle.  2) The influence of the thermal fluctuations on the
kinetic process was examined in the framework of the nearest-neighbor
interaction model. The thermal fluctuations lead to the additional
splitting of the steps on the magnetization curve. 3) The magnetization
curve for system with interaction and coercivity dispersion was
calculated in mean field approximation.  The simple method to
experimentally distinguish the influence of interaction and coercivity
dispersion on the magnetization curve is suggested.

\section{Introduction}
The properties of the magnetic nanoobjects and their systems are of
current concern due to the appearance of technological possibilities of
their fabrication and measurements. The reason of such interest is that
such systems are ideal for studying collective effects and phase
transitions and are also attractive as a media for high-density magnetic
storage. Today there are the experimental data for the magnetization
curves of the 2D quadratic lattice of the Ni pillars with the
perpendicular anisotropy \cite{hwang_00,farhoud_00}, for the rectangular
\cite{gusev,cowburn_NJF} and square \cite{cowburn_99,mathieu_97} lattices
of the permalloy nanoparticles, for the square lattice of CoCrPt
particles with the perpendicular anisotropy \cite{haginoya_99} for the
system of the Fe \cite{hauschild_98} and permalloy \cite{adeyeye_97}
nanostripes, for the 2D systems of the chains of the Co particles
\cite{martin_98}, for anisotropic linear self-assembling mesoscopic Fe
particle arrays \cite{sugawara_97_1,sugawara_97_2}.

What is the main common feature of represented magnetic systems from the
theoretical point of view? They consists of the magnetic coercive
objects: magnetic particles with the perpendicular single-particle
anisotropy \cite{hwang_00,farhoud_00,haginoya_99}; particle chains, which
have the effective anisotropy axis along the chain due to interparticle
dipole-dipole interaction
\cite{gusev,martin_98,sugawara_97_1,sugawara_97_2}; magnetic nanostripes
having the form anisotropy \cite{hauschild_98,adeyeye_97}. In the
individual magnetic nanostripe the magnetization process take place by
nucleation-propagation process \cite{zhukov_97,pignard_00}. The
propagation of the nucleus is very rapid. The numerical simulation
demonstrates, that in 1D nanoparticle chain the magnetization reversal
proceeds through nucleation and the following domain wall propagation
also \cite{mism}. Both magnetic nanostripe and magnetic nanoparticle
chain have two stable states within the external magnetic field with the
magnetization directed along the stripe or the chain. A nanoparticle with
the perpendicular anisotropy axis also has two stable states.  The
magnetization reversal in the particle has thermoactivated character
\cite{braun_93}. In the system of such magnetic nanoobjects there is the
long-range interaction between them which have the magnetostatic nature.
In systems of the magnetic particles with perpendicular to the array
plain anisotropy the interaction has the effective antiferromagnetic
character.  Its energy

\begin{equation}
E_{ij}=\frac{M(r_{i})M(r_{j})}{|r_{ij}|^3},
\label{eperp}
\end{equation}

\noindent
$M(r_{i})$ is a particle magnetic moment, $r_{ij}$ is interparticle
distance.  In the case of the system of the magnetic nanostripes the
magnetostatic interaction are of long-range character also. It caused by
magnetic charges appeared on the stripe edges in the magnetized state.
The dependence of the interaction energy on the inter stripe distance is

\begin{equation}
E_{stripe}=2\frac{M(r_{i})M(r_{j})}{L^2}
(\frac{1}{r_{ij}}-\frac{1}{\sqrt{r_{ij}^2+L^2}}),
\label{estr}
\end{equation}

\noindent
here $M(r_{i})$ is stripe magnetic moment, $r_{i,j}$ is inter stripe
distance, $L$ is their length.  For the neighboring stripes $E \sim
r^{-1}$, on the long distances $E \sim r^{-3}$.

Another system is 2D rectangular lattice of the magnetic nanoparticles
with the single-particle anisotropy of the "easy plain" type. In this
case particles form chains lying along the short side of the elementary
rectangular. Due to anisotropy character of the dipole interaction the
chain magnetization directed along the chain.  The energy of the inter
chain interaction involves the part (\ref{estr}) connected with the
existence of the magnetic charges on the chain edges and the part caused
by discreteness nature of the chain \cite{gross_97,rozenbaum_91r}:
\begin{equation}
E_{discr}=\frac{8\pi^2 M(r_{i})M(r_{j})/La^2}
{\sqrt{r_{ij}/a}}exp(-2\pi r_{ij}/a),
\label{edir}
\end{equation}

\noindent
here $M(r_{i})$ is magnetic moment of a chain, $a$ is the interparticle
(within chain) distance, $r_{ij}$ is the distance between the chains and
$L$ is their length. The relation $E_{discr}/E_{strip}$ is proportional
to the chain length. So for the sufficiently long chains the
nearest-neighbor interaction plays a leading role \cite{gross_97}. The
character of interaction is antiferromagnetic too.

There were made some attempts to solve the problem of the magnetization
process in the coercive system with the interaction by numerical methods
\cite{grundler_99,gonzalez_98,brown_00}. It was found, that the
magnetization curves look like staircase with the steps of different
widths. Besides there is the first report about the experimental
observation of such steps \cite{grundler_99}.

In our work we provide analytical investigation of the problem. In the
first part we solve 1D model of system of $2^N$ magnetic moments with
coercivity and long-range interaction decaying proportionally $1/r^p$
which corresponds to the system of magnetic nanostripes. We use the
cyclic boundary conditions. It was shown that the magnetization curve
consists of the series of steps corresponding to formation of
superstructures. In the case $N\rightarrow\infty$ this staircase-like
curve has self-similar character (so-called "devil-staircase"). The
method of the solution can be easily generalized on the case of 2D
quadratic lattice of the magnetic nanoparticles with perpendicular
anisotropy. In second part we solve the problem for 1D system with
niarest-neighbor antiferromagnetic interaction and single object
coercivity at finite temperature. The influence of thermal fluctuation
on the magnetization process described in first part is also discussed
here. It is shown that the appearance of defects in the superstructures
leads to splitting of the steps on the magnetization curve. The influence
of the coercivity dispersion is also discussed. In the third part of the
article we use the mean-field approximation to show, how one can
distinguish interaction and coercivity dispersion in the system of
magnetic nanoobjects. Very easy method to estimate the contribution of
interaction and coercivity dispersion in the magnetic properties of the
system is suggested.

\section{The model with the long range interaction.\\
"Devil-staircase"}

Let us consider the 1D system of long-range interacting coercive magnetic
objects.  Its appropriate example is system of finite length magnetic
nanostripes or chains of magnetic nanoparticles. There is the effective
antiferromagnetic interaction of magnetostatic nature between them. We
consider its energy in the dimensionless form

\begin{equation}
\epsilon=\frac{I}{|k-n|^p}\sigma_k\sigma_n,
\label{J}
\end{equation}

\noindent
here $\sigma_k=\pm1$ are interacting magnetic moments, $n$ and $k$ are
the numbers of magnetic moments positions, $I$ is a dimensionless
constant of the effective antiferromagnetic interaction ($I>0$). The
nearest-neighbor object distance is equal to 1.  Let the system is
magnetized so, that all $\sigma_k=-1$. The magnetization reversal in the
totally magnetised system onsets then the field at the object place
exceeds its coercivity.  In the case of the infinite chain it is

\begin{equation}
H_1=H_c-2I\xi(p), \qquad   \xi(p)=\sum_{k=1}^{N}\frac{1}{k^p}.
\label{begin}
\end{equation}

\noindent
Here $H_c$ is coercivity of the object and it is the same for all
objects. The second term is a field originating at the place of one
object by all the others.

\noindent
In a similar way it can be easily calculated, what reversal ends at the
external field value

\begin{equation}
H_2=H_c+2I\xi(p),
\label{finish}
\end{equation}

\noindent
when the last object will be reversed. It is interesting to investigate
magnetization curve in the region $H_1<H<H_2$. As the system is
one-dimensional, its ground state is disordered. Here we will consider
the temperatures less than single object coercivity ($kT<MH_c$), so the
system can be in a number of metastable states. For example, if the
interaction energy in the system is less than energy of coercivity there
are $2^n$ (n is a number of objects) metastable states at zero external
field.

If objects do not have coercivity or temperature of the system is rather
high ($kT>>MH_c$) system will be in a ground state at any moment. If the
external magnetic field is less than $-2I\xi(p)$ or larger then
$2I\xi(p)$ the ground state is totally magnetised state. The values of
the magnetization of the ground states corresponding to external field at
the interval $-2I\xi(p)<H<2I\xi(p)$ can be also found. This
thermodynamical problem was solved in \cite{bak_82}. It was obtained that
in this case magnetisation curve has steps and looks like self similar
"devil-staircase".

Here we solve the problem for the case when $kT$ is less than any energy
in the system. So system can be in metastable states and the problem of
the magnetization reversal can not be solved by thermodynamical methods.
We must solve the kinetic problem by correct choice of a consequence of
metastable states the system passes through. The problem of such choice
become easier in the case when the system consicts of absolutely even
number ($N=2^n$) of objects and have cyclic boundary conditions.

So the magnetization reversal onsets then the external field exceeds
$H_1$.  Fluctuations are the reason that the place of the first reversed
object can be choused arbitrary. Due to antiferromagnetic character of
the interaction the additional increase of the external field is
necessary to reverse second object. As the interaction decreases with the
inter object distance, the second reversed object must be chosen at
the largest distance from the first one. It is very easy to choose this
place in the case of the system with cyclic boundary conditions (Fig.
1,I).

\begin{figure}[th]
\centerline{
\epsfxsize=7cm
\epsffile{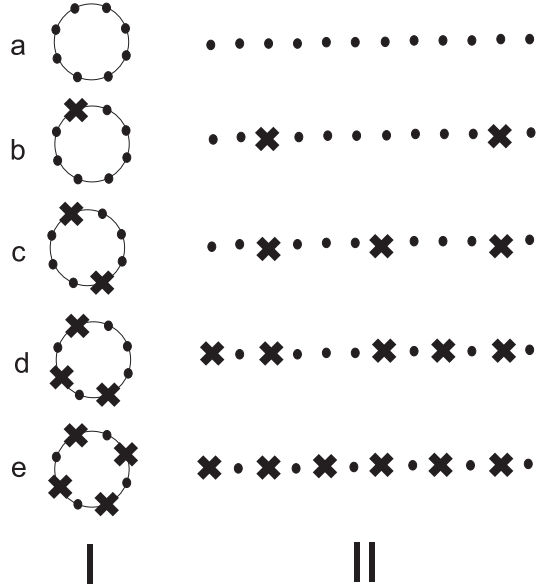}}
\caption[b]{
The reversal process in the cyclic system of 8 magnetic moments and
corresponding superstructures in the infinite system. Points mark not
reversed magnetic moments, crosses mark already reversed ones. The
external field increases from a) to e). I) the reversal process in the
cyclic system of 8 objects, II) the corresponding superstructures in the
infinite system.
} \end{figure}

\noindent
 In this case the magnetization process takes place through the
sequential formation of different superstructures (Fig. 1,II) which are
metastable states.  Let us calculate the field values, then the definite
superstructure appears ($H_-$) and became unstable ($H_+$). At first we
consider only superstructures with period $m=2^k$, $k=0,1,2...$ and one
reversed object per period (such in Fig. 1 b) c) and e)), $k=0$
corresponds to fully magnetized state.  The field, when the
superstructure with the period equal to $m$ is formed (Fig. 2), is

\begin{equation}
H_-(m)=H_c+2I\frac{\xi(p)}{m^p}-(2I\xi(p)-2I\frac{\xi(p)}{m^p})
=H_c-2I\xi(p)+4I\frac{\xi(p)}{m^p}.
\label{minus}
\end{equation}

\begin{figure}[th]
\centerline{
\epsfxsize=7cm
\epsffile{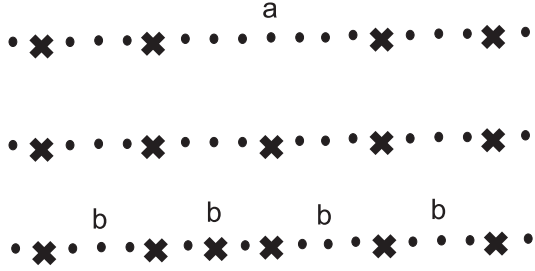}}
\caption[b]{
\noindent
The process of the formation and following destruction of the
superstructure with the period $m=4$. Points mark not reversed magnetic
moments, crosses mark already reversed ones. $H_-$ is the field when the
magnetic moment in the point (a) becomes unstable and the formation of
the superstructure takes place. $H_+$ is the field when the magnetic
moments in on of the points (b) become unstable. Due to fluctuations one
of them reverses first of all and then prevent others to reverse due to
long-range antiferromagnetic interaction.
} \end{figure}

\noindent
Here the second term is the fields of already reversed objects. They
prevent the chosen object to reverse. The term in brackets is the field
of the other not reversed objects. They help chosen object to reverse. In
a similar way, the field then the superstructure with the period equal to
$m$ become unstable (Fig.2) is

\begin{equation}
H_+(m)=H_c
+2I\frac{(2^p-1)\xi(p)}{m^p}-(2I\xi(p)-2I\frac{(2^p-1)\xi(p)}{m^p})
\label{plus}
\end{equation}
$$
=H_c-2I\xi(p)+4I\frac{(2^p-1)\xi(p)}{m^p}.
$$

\noindent
Here we use the relation

\begin{equation}
\sum_{n=1}^{\infty}\frac{1}{(n/2)^p}=
\sum_{n=1}^{\infty}\frac{1}{(n+1/2)^p}+\sum_{n=1}^{\infty}\frac{1}{n^p},
\end{equation}

\noindent
The magnetization of the system is determined by the superstructure period
and is

\begin{equation}
M=\lim_{N\to\infty}\frac{1}{2^N}\sum_{k=1}^{\infty}\sigma_k=\frac{2-m}{m}
\label{m}
\end{equation}

\noindent
So there are the mast be steps on the magnetization curves corresponding
to the stable superstructures, as the magnetization does not change while
the magnetic field changes from $H_-(m)$ to $H_+(m)$. Using
(\ref{minus},\ref{plus},\ref{m}) we can rewrite the dependence of the
step edges of the magnetization (Fig. 3) in the form:

\begin{equation}
H_-=H_c-2I\xi(p)+4I\xi(p)(\frac{M+1}{2})^p,
\label{minus_M}
\end{equation}
\begin{equation}
H_+=H_c-2I\xi(p)+4I(2^p-1)\xi(p)(\frac{M+1}{2})^p,
\label{plus_M}
\end{equation}

\begin{figure}[th]
\centerline{
\epsfxsize=7cm
\epsffile{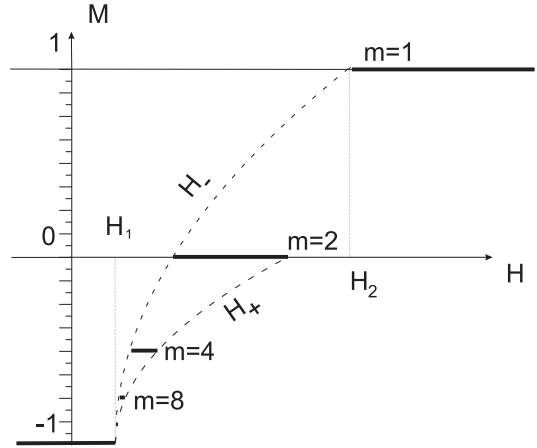}}
\caption[b]{
\noindent
The steps on the magnetization curve corresponding to the simple
superstructures ($m=2^k$) with one reversed object per period.
Magnetization of the corresponding step is $M=(2-m)/m=-1+1/2^{k-1}$
} \end{figure}

\noindent
The steps corresponds to the reviewed superstructures do not cover all
fields between values $H_1$ and $H_2$ (Fig. 3). To analyze the
magnetization behavior of the system while it transfers from one step to
another we must take into account the formation of the more complex
superstructures (Fig. 1d).  Let us consider, for example, the
magnetization process between $H_+(m=2)$ and $H_2$ (Fig. 3), that is how
antiferromagnetic superstructure becomes fully magnetized. The
superstructures with periods $m_2=2,4...2^l$ ($l=1,2...$) appearing with
antiferromagnetic ($m_1=2$) as a background are represented at Fig.4.

\begin{figure}[th]
\centerline{
\epsfxsize=7cm
\epsffile{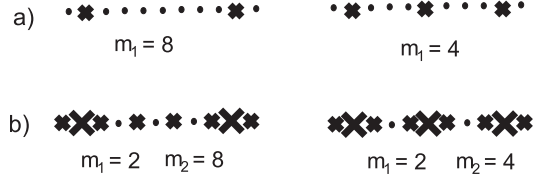}}
\caption[b]{
a) are the simple superstructures ($m_1$ is a period). Points
mark not reversed magnetic moments, crosses mark already reversed ones.
b) are complex superstructures characterized by two numbers $m_1$ and
$m_2$.  Small crosses mark antiferromagnetic background, big crosses mark
reversed magnetic moments forming period of a superstructure.
} \end{figure}

\noindent
$m_1=2, m_2=2$ is totally magnetized state, $m_1=2, m_2=\infty$ is
antiferromagnetic superstructure. In this case $M=2/m_2$. The expressions
for the magnetic fields $H_-(m_2)$ and $H_+(m_2)$ differ from the case of
the simple superstructures as we must take into account the field of the
antiferromagnetic background affecting on the objects.  This additional
field is

\begin{equation}
H=2I\frac{(2^p-1)\xi(p)}{m_1^p}=2I\frac{(2^p-1)\xi(p)}{2^p}
\end{equation}

\noindent
We must take this field into account twice as formerly it was directed
along the external field but now it is directed against it. So we have

\begin{equation}
H_-(m_2)=H_c-2I\xi(p)+4I\xi(p)(\frac{M}{2})^p
+4I\frac{(2^p-1)\xi(p)}{2^p}
\label{minus_M2}
\end{equation}
\begin{equation}
H_+(m_2)=H_c-2I\xi(p)+4I(2^p-1)\xi(p)(\frac{M}{2})^p
+4I\frac{(2^p-1)\xi(p)}{2^p}
\label{plus_M2}
\end{equation}

\noindent
Evidently (\ref{minus_M2},\ref{plus_M2}) are exact similar to
(\ref{minus_M},\ref{plus_M}), but this curves starts at the point $M=0$,
$H=H_+(m_1=2)=H_c-2I\xi(p)+4I\frac{(2^p-1)\xi(p)}{2^p}$, which is the
right edge of the step corresponding to the antiferromagnetic structure,
instead of $M=-1$, $H=H_1=H_c-2I\xi(p)$. In a similar way all other steps
on the magnetization curve can be obtained. Each step corresponding to
the superstructure is the base of a series of the steps, corresponding to
the more complex superstructures with the first one as a background. The
dependencies of $H_-$ and $H_+$ in all cases are similar, and $H_\pm\sim
M^p$. So the picture become self-similar (Fig. 5).

\begin{figure}[th]
\centerline{
\epsfxsize=7cm
\epsffile{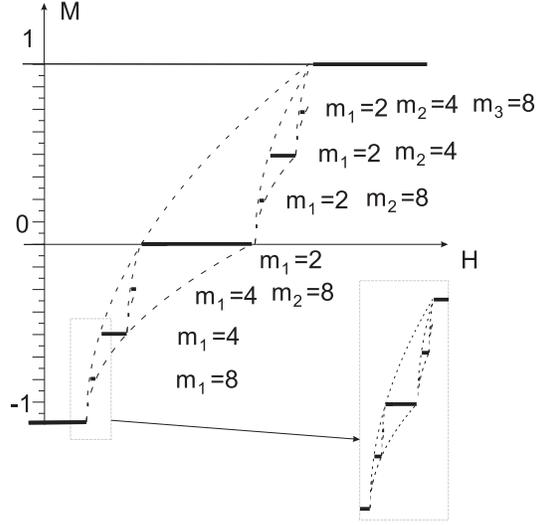}}
\caption[b]{
The self-similar "devil staircase".
} \end{figure}

\noindent
Complex superstructures are characterized by a series of numbers
$m_1,m_2,m_3...$, where $m_i=2^k$ and $m_i<m_{i+1}$. Maximal $m_{max}$ is
the period of superstructure, $m_1,m_2 ... m_{max-1}$ characterizes its
background. Narrow steps are between wider ones.  Let as find the common
width of all steps. The step width can be easily calculated.

\begin{equation}
\Delta H=H_+(m)-H_-(m)=4I\frac{2^p-2}{m_{max}^p}\xi(p)
\end{equation}

\noindent
the number of the steps with equal width depends on $m_{max}$ and
$N=m_{max}/2$. The common width is

\begin{equation}
\Delta H=\sum_{\mbox{n is even}}^{\infty} N(n)\Delta H(n)
=\sum_{\mbox{n is even}}^{\infty} \frac{n}{2}4I\frac{2^p-2}{n^p}\xi(p)
=4I\xi(p)\sum_{k=1}^\infty 2^{k-1}\frac{2^p-2}{2^{kp}}
\end{equation}
$$
=2I\xi(p)(2^p-2)\sum_{k=1}^\infty (\frac{1}{2^{p-1}})^k
=2I\xi(p)(2^p-2)\frac{(1/2)^{p-1}}{1-(1/2)^{p-1}}=4I\xi(p)
$$

\noindent
So the sum of all steps widths is exactly the same as the width of the
whole inclination of the magnetization curve, which is equal to
$4I\xi(p)$ according to (\ref{begin},\ref{finish}).

The difference in the step width is connected with the decaying character
of the long-range interaction. The wider ones are conditioned by
interaction of more near magnetic moments; the narrow ones are caused by
the interaction of more distant magnetic moments.

The magnetization values corresponding to the steps are

\begin{equation}
M=\frac{4m-1}{m_{max}}-1,
\end{equation}

where $m=1,2...m_{max}/2$.

It is interesting, that the exponent in (\ref{minus_M}) and
(\ref{plus_M}) is equal to the power index in the dependence of the
interaction on the inter object distance (\ref{J}). So it is possible to
find the power index for the interaction by the experimental measuring of
magnetization curves.

There are some reasons of the distortion of the represented ideal
picture. At first it is thermal fluctuations. As for the long distances
($kT>E_{int}(r)$, $k$ is a Boltzman constant) it leads to fuzzifying and
disappearing of narrow steps which are conditioned by interaction of
distant magnetic moments.  Besides thermal fluctuations can lead to
appearance of the defects in superstructures.  This will be discussed in
the next section.  The second reason is the bounds of the real system.
They can play a significant role as the interaction has a long-range
character.  Nevertheless if the dimension of the system is larger than
$r_0$ ($kT=E_{int}(r_0)$), the influence of the bounds will be neglected
by thermal fluctuations. Lastly the dispersion of the objects coercivity
can dramatically change the magnetization curve. Such self-similar
behavior can be observed only in the system with small (less than
interaction) coercivity dispersion. The method how to distinguish the
influence of the interaction and coercivity dispersion in possible
experiment is discussed in the last section.

In spite of all deficiencies of the proposed model it help to understand
the peculiarities of the magnetization process in the system, the nature
of the steps on the magnetization curve
\cite{grundler_99,brown_00,sampaio} and especially the fact that the
difference in steps width is a sequence of the decaying character long
range interaction. It also makes understandable the fact of the
alternation of the narrow and wide steps on the magnetization curve
\cite{grundler_99,brown_00}. It is very probably that in the general case
the magnetization curve has self-similar character too. The model can be
easily generalized on the case of the 2D square lattices of the magnetic
nanoparticles. In this case one must provide the summation of the dipole
sums for the superstructures on the square lattice. The superstructures
must have square elementary cell in this case, so the dipole sums can be
easy calculated \cite{jmmm}.

\section{The nearest-neighbor model.\\
Thermal fluctuations}

Let us consider the magnetization process at finite temperature less than
the coercive energy of a single magnetic moment $kT<H_c M$ but higher
than interaction energy of distanced magnetic moments.  In this case
magnetic moments at distances larger than $r_0$  ($kT=E(r_0)$) begin to
reverse independently as their interaction is smaller than the
temperature.  Nethertheless, as $kT<H_c M$ system can be in metastable
states.  The kinetics of magnetization process can be qualitatively
described as follows. When the external magnetic field exceeds the value
of $H_1$ the thermoactivated reversal of the individual magnetic moments
begins. But already reversed magnetic moments prevents neighboring
magnetic moments (lying at distances smaller than $r_0$) to reverse due
to effective antiferromagnetic interaction. More distant magnetic moments
can reverse, as the interaction energy in this case is smaller than
temperature.  Magnetization process has Poisson-like character and ends
when the distance between reversed magnetic moments will be in the
interval $r_0<r<2r_0$. The addition external field is necessary to
overcome antiferromagnetic interaction and to continue the process of the
reverse.  Evidently the thermal fluctuations will lead to distortion of
the ideal picture described in the previous section. It is difficult to
take thermal fluctuations in to account in the general case. Here we have
solved the problem for the situation when the temperature is large than
any interaction in the system with exception of the most powerful
nearest-neighbor interaction.  In this case the problem can be solved in
the nearest-neighbor approximation ($p=\infty$ in (\ref{J})). This model
is also appropriate for the case of the planar system of the long chains
of the magnetic nanoparticles when the main term in the interaction is
interaction of nearest neighbors (\ref{edir}).  The form of the
hysteresis in this case is represented in Fig.6. The magnetization
reversal starts at the field value $H=H_c-2I$, as the antiferromagnetic
interaction helps the external field.

\begin{figure}[th]
\centerline{
\epsfxsize=7cm
\epsffile{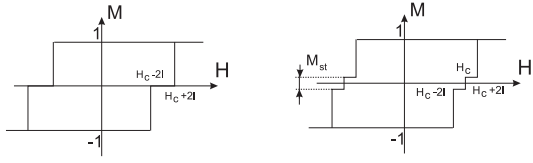}}
\caption[b]{
The hysteresis loop in the case of the nearest-neighbor interaction.
a) in the case of $T=0$ the step corresponds to antiferromagnetically
ordered superstructure.
b) the dividing of step in two due to formation of defects in the case of
thermal fluctuations. $M_{st}$ is a step high.
} \end{figure}

\noindent
The chains begin to reverse the
magnetization due to thermal fluctuations. But if the one chain reverses,
it prevents the neighboring chains to reverse as the effective field
of the interaction is opposite to external field.  Due to chaotic
character of the sequence of the magnetization reversals of the elements
of the system the defects appearances is possible (Fig. 7), and the
antiferromagnetic ordering with $M=0$ does not be achieved at this value
of the external field.  The additional external field ($H=H_c$) is
necessary to reverse the defects. Then defects change their sign (Fig.7).

\begin{figure}[th]
\centerline{
\epsfxsize=7cm
\epsffile{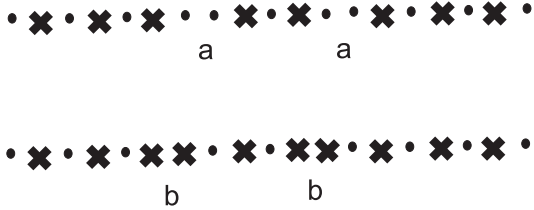}}
\caption[b]{
(a) are defects appearing due to Poisson-like character of the
magnetization process, (b)  are reversed defects. Points mark not
reversed magnetic moments, crosses mark already reversed ones.
} \end{figure}

\noindent
The magnetization reversal ends within the field $H=H_c+2I$ when system
becomes totally magnetised.  So two steps appear on the each branch of
the hysteresis loop.  Their width is $\Delta H=2I$ i.e. it depends on the
interaction value.  The magnetization value $M_{st}$ corresponding to the
step depends of defects concentration. It is the special problem to find
this concentration. Let us consider the kinetics of the defects
appearance in the system of $N$ objects. Firstly all objects are
magnetized against external magnetic field $\sigma_k=-1$.  When the
reversal process begins, the reversed objects begin to divide system
into the regions of the not reversed objects (we will name them as
"domains"). In process of magnetization the number of domains increases,
their widths decrease. Let $P_n$ is a number of domains consisted of $n$
not reversed objects then

\begin{equation}
M=\sum_{n=1}^{N-2}P_n-\sum_{n=1}^{N-2}nP_n,
\end{equation}

\noindent
If $\alpha$ is the probability of the chain reverse per unit time, then

\begin{equation}
\frac{\partial P_n}{\partial t}=-\alpha (n-2)P_n+2\alpha\sum_{k=n+2}^N
P_k, \qquad n>2.
\label{6323}
\end{equation}

\noindent
The first term is decrease of the domain number due to its dividing into
smaller ones; the second term describes the appearance of new domains due
to dividing of wider ones. As the domains consisted of one or two objects
can not further divide, their number increases only. So

\begin{equation}
\frac{\partial P_n}{\partial t}=2\alpha\sum_{k=n+2}^N P_k,
\qquad n=1,2.
\label{6324}
\end{equation}

\noindent
Evidently $\alpha$ depends on temperature, but as we in the states stable
at $t\rightarrow\infty$, $\alpha$ does not affect final result. It can be
easily checked, that

\begin{equation}
\frac{\partial}{\partial t}\sum_{n=1}^N (n+1)P_n=0,
\end{equation}

\noindent
i.e. the whole number of objects in the system is constant. In the course
of time ($\alpha t>>1$) only the domains with $n=1,2$ remains. So
$P_n(t\rightarrow\infty)=0$ for $n>2$ and $M=-P_2(t\to\infty)$. Let us
use the Laplas transformation and define

\begin{equation}
p_n(s)=\int_0^\infty P_n(t)e^{-st}dt.
\end{equation}

\noindent
Then, according to (\ref{6323})

\begin{equation}
p_n(s)=\frac{2\alpha}{\alpha(n-2)+s}Q_n(s),
\end{equation}

\noindent
where

\begin{equation}
Q_n(s)=\sum_{k=n+2}^{N}p_k(s).
\label{Q}
\end{equation}

\noindent
For $s\rightarrow 0$

\begin{equation}
Q_n(0)=\sum_{k=n+2}^{N}p_k(0)=\sum_{k=n+2}^{N}\frac{2}{k-2}Q_k(0).
\end{equation}

\noindent
In the recursive form this equation can be written as

\begin{equation}
Q_n(0)=Q_{n+1}(0)+\frac{2}{n}Q_{n+2}(0)
\label{rec}
\end{equation}

\noindent
To find the magnetization value corresponding to the step it is necessary
to calculate $P_2(t\rightarrow\infty)$. Evidently

\begin{equation}
P_2(t\rightarrow\infty)=\lim_{s\to 0}sp_2(s)=2\alpha Q_2(0)
\end{equation}

\noindent
If the maximum number of the objects in the system is $N$, $Q_2(0)$ can
be found from the (\ref{rec}) with the initial conditions
$Q_{N-2}=Q_{N-3}=p_N(0)$, which are the sequence of (\ref{Q}). In its
turn $p_N(0)=1/\alpha(N-2)$ according to (\ref{6323}). The solution of
the recursive equation was found numerically as

\begin{equation}
M_{st}=2\alpha \lim_{N\to\infty}\frac{Q_2 (0,N)}{N} \approx 0.134
\end{equation}

\noindent
As $Q_2 (0,N)$ is proportional to $\alpha^{-1}$, the result does not
depend on the value of $\alpha$. So the defects formation during
magnetization leads to appearance of two steps on the magnetization curve
in the case of the nearest-neighbor interaction. The fluctuational
character of the magnetization process prevent chance to find system in
the antiferromagnetically ordered ground (if $H=0$) state. One can expect
in the case of the long range interaction thermal fluctuations will lead
to similar splitting of the magnetization steps too.

\section{Dispersion of the coercivity}

Another reason of the distortion of the described magnetization process
is the dispersion of the coercivity in the system. It is necessary to
distinguish the effects of the interaction and the coercivity dispersion
especially in experiment. This problem is solved here in the mean field
approximation.  In the framework of this model the interaction is
independent on the distance and $\epsilon=I/N$. It must be mentioned that
this model is appropriate in the case of the very long stripes when
$L>Na$ ($L$ is a stripe length, $a$ is interstripe distance).  The
dependence of the magnetization on the external field is linear in this
case as

\begin{equation}
H=H_c+JM
\end{equation}

\noindent
$H_c$ is coercivity, $JM$ is the field of interaction in the mean field
approximation. The main feature of the magnetization process in coercive
system with the antiferromagnetic interaction is connected with its
multistability. It means that if we change the sign of the external field
changing the system does not change its magnetization immediately.  At
first it transits through the whole hysteresis loop from one branch to
another (from point A to point B, Fig. 8),

\begin{figure}[th]
\centerline{
\epsfxsize=7cm
\epsffile{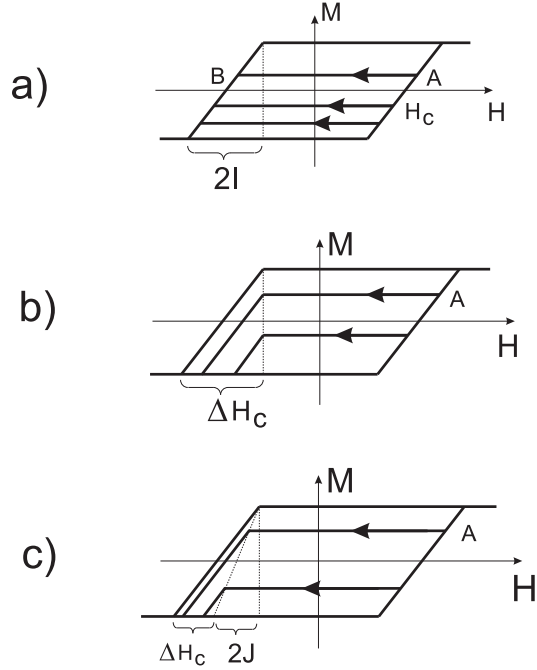}}
\caption[b]{
The hysteresis curve in the mean field approximation. a) in the case of
the system with interaction. b) in the case of the coercivity dispersion
(uniform distribution of the coercivity). c)system both with interaction
and coercivity dispersion.
} \end{figure}

\noindent
and then begins to change
according to the new branch. In this case $|H_A-H_B|=2H_c$.  It must
be noted that such behavior does not depend on the interaction manner. So
different states of the system (characterized by the different
magnetization) correspond to the same value of the external field. Such
type of the multistability we will name interaction-type (I-type),
because the multistability can be in the system of the non-interacting
magnetic objects with different values of coercivity also. In this case
magnetization reversal begins when field reaches the value
$H_1=H_{Cmin}$, when the reversal of the objects with the smallest
coercivity starts. The reversal process is finished at the field value
$H_2=H_{Cmax}$, corresponded to the largest coercivity in the system.
The hysteresis loop in this case the similar to the one for the system
with interaction (as its branches can have similar inclination in the
case $\Delta H_c=2I$ and uniform distribution of $H_c$), but the
transitions inside the loop differs.  If one change the direction of the
reversal process in the A point (Fig.8) in this case, the magnetization
does not change while the external field reaches the value of $-H_1$ when
the reversal of the objects with the smallest coercivity happens. This
type of the multistable behavior we will name coercivity-type (C-type).
The hysteresis loop of the system both with the interaction and
coercivity dispersion can be easily calculated too.  It is represented on
Fig.8c. So one can distinguish interaction and coercivity dispersion by
the behavior of the magnetization within hysteresis loop by analysis of

 the multistability type of the system.  Evidently the self-similar
behavior of the magnetization can be experimentally observed in the
systems with small dispersion of coercivity, that is in the systems which
demonstrate I-type of multistability.   Let us note, that in the reviewed
experimental works the type of the multistability is not examined in
spite of the simplicity, from the on hand, and, significance, from the
other hand, of such investigation.

\section{Conclusions}

By means of the simple models we have investigated the magnetization
processes in the systems of the coercive magnetic objects with
interaction. The reason of the formation of the steps on the
magnetization curve is investigated. It is shown that the magnetization
curve can have self-similar character. Its form is calculated in the case
of the long-range interaction with $E\sim 1/r^p$. The influence of the
thermal fluctuations is analyzed in the framework of the nearest-neighbor
approximation. It is shown that fluctuations lead to splitting of the
steps on the magnetization curve. The effects of the interaction and
coercivity dispersion on the hysteresis loop are examined by mean field
approximation.  Their difference is shown. It consists in the character
of the magnetization dependence on the external magnetic field during the
transition between the branches of the hysteresis loop. The understanding
of this difference is very important to interpret of the experimental
data.  It allows to experimentally distinguish the influence of the
coercivity dispersion and interaction in the system on the magnetization
processes.

$\qquad$ \section*{Acknowledgment.}
We are grateful to prof. A.A.Andronov for helpful discussion. The work
was supported by the Russian Foundation for Fundamental Research (N
00-02-16485).  \bibliography{dipole} \bibliographystyle{ms_phrev}

\end{document}